# Hafnium carbide formation in oxygen deficient hafnium oxide thin films

C. Rodenbücher,[1] E. Hildebrandt,[2] K. Szot,[1,3] S. U. Sharath,[2] J. Kurian,[2] P. Komissinskiy,[2] U. Breuer,[4] R. Waser,[1,5] and L. Alff [2]

[1] Forschungszentrum Jülich GmbH, Peter Grünberg Institute (PGI-7), JARA-FIT, 52425 Jülich, Germany
[2] Technische Universität Darmstadt, Institute of Materials Science, 64287 Darmstadt, Germany
[3] University of Silesia, A. Chełkowski Institute of Physics, 40-007 Katowice, Poland
[4] Forschungszentrum Jülich GmbH, Central Institute for Engineering, Electronics and Analytics (ZEA-3), 52425 Jülich, Germany
[5] RWTH Aachen, Institute of Electronic Materials (IWE 2), 52056 Aachen, Germany

On highly oxygen deficient thin films of hafnium oxide (hafnia, $HfO_{2-x}$) contaminated with adsorbates of carbon oxides the formation of hafnium carbide ($HfC_x$) at the surface during vacuum annealing at temperatures as low as 600 °C is reported. Using X-ray photoelectron spectroscopy the evolution of the $HfC_x$ surface layer related to a transformation from insulating into metallic state is monitored in situ. In contrast, for fully stoichiometric $HfO_2$ thin films prepared and measured under identical conditions, the formation of $HfC_x$ was not detectable suggesting that the enhanced adsorption of carbon oxides on oxygen deficient films provides a carbon source for the carbide formation. This shows that a high concentration of oxygen vacancies in carbon contaminated hafnia lowers considerably the formation energy of hafnium carbide. Thus the presence of a sufficient amount of residual carbon in resistive random access memory devices might lead to a similar carbide formation within the conducting filaments due to Joule heating.

$HfO_2$ (hafnia) has been studied extensively in the last decades as a high-$\kappa$ dielectric being used as replacement of the standard gate dielectric $SiO_2$ in order to produce high-density logic and memory devices.[1] Furthermore, it was found that $HfO_2$ can undergo a local insulator-metal (IM) transition under the influence of an electric field opening up the possibility to use it as resistive switching material in redox-based random access memories (RRAM).[2,3]

During resistive switching in $HfO_2$, a local redox reaction associated with a formation or modification of conducting filaments takes place, in which the movement of oxygen vacancies plays the key role for the modulation of the electric properties.[4–6] Hence, the investigation of the behavior of oxygen vacancies is important for RRAM device performance.

In contrast to the local insulator-metal transition as observed in conducting filaments in $HfO_2$ on the nanoscale, also a global insulator-metal transition on a macroscopic scale has been found in oxygen deficient hafnium oxide thin films grown by molecular beam epitaxy (MBE).[7,8] The oxygen deficiency in these films is created during growth under oxygen deficient conditions leading to a homogeneous distribution of oxygen vacancies. With increasing amount of oxygen vacancies, the band gap is consistently reduced and defect states inside the gap eventually form a defect band at the Fermi level.[7,8] RRAM devices based on these films show forming free resistive switching and the forming voltage is independent on the thickness of the highly oxygen deficient $HfO_{2-x}$ layer.[9,10]

In this paper we discuss the reaction of carbon impurities to hafnium carbide and its influence on hafnium oxide-based resistive switching. The role of carbon impurities is of extremely high importance, as they are inherent to most of today's deposition techniques commonly used for fabricating complementary metal oxide semiconductor (CMOS) devices, such as atomic layer deposition (ALD) based on organic precursors. Based on an *in-operando* hard X-ray photoelectron spectroscopy study, it was concluded that carbon impurities are detrimental to the cell endurance[11]. In contrast to that, it was suggested that intentional doping with carbon could also be exploited in order to enhance the performance of resistive switches.[12] We study the effect of a controlled thermal treatment under ultra-high vacuum (UHV) conditions on the chemical reactions of residual carbon, both, for stoichiometric and oxygen deficient films of hafnium oxide. Studying the effect of post annealing on carbon impurities is important, as during CMOS processing several heat treatments of the whole device are performed, and the sample is locally heated during multiple resistive switching.

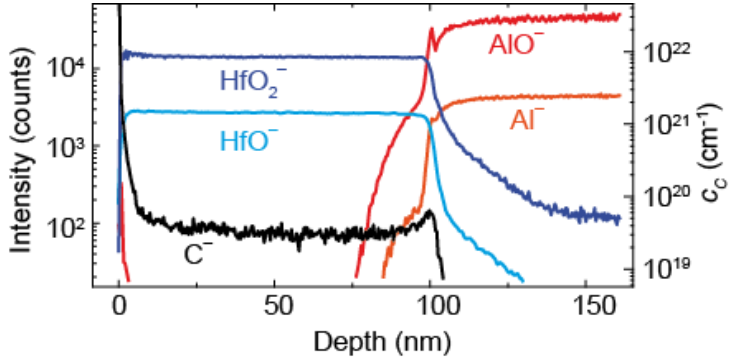

Fig. 1: ToF-SIMS depth profile of a 100 nm thick HfO$_{2-x}$ film deposited on Al$_2$O$_3$. The carbon concentration $c_C$ was calculated using a carbon-implanted calibration sample.

HfO$_{2-x}$ thin films with thicknesses of 100 Å were grown by MBE on $c$-cut sapphire substrates under various oxygen flow rates of 0.3, 0.5, 1.0, and 1.5 standard cubic centimeters per minute (sccm) corresponding to highly oxygen deficient to stoichiometric, while all other deposition parameters were kept constant. A detailed description of the deposition parameters can be found in Ref. 7. In this way, a series of oxygen deficient hafnium oxide thin films with various oxygen vacancy concentrations was synthesized. After growth, the electronic structure and chemical composition of the films were investigated by XPS using a Physical Electronics instrument with monochromatized Al-K$_\alpha$ rays in UHV conditions after *ex situ* sample transfer. The angle of analysis was set to 45° resulting in an information depth of about 6 nm. In order to avoid charging of the sample during measurement, electron irradiation was used to neutralize the surface. All samples were measured before, during, and after *in situ* annealing inside the XPS chamber in order to study the electronic structure as a function of annealing time and temperature under UHV conditions. The recorded XPS spectra were simulated by peak fitting after background subtraction using the Shirley method. The intrinsic carbon content of the films was investigated by time-of-flight secondary ion mass spectrometry (Tof-SIMS). A known amount of carbon was implanted into one film using an Eaton implanter (11 keV, 36 min, 24 µA) allowing for Tof-SIMS intensity calibration. LC-AFM as a surface sensitive tool for the local investigation of the nature of the resistive switching effect in oxides has been applied to obtain information on film morphology and local electrical conductivity.[13,14] For this purpose, a JEOL-AFM system equipped with Pt/Ir coated tips was used in contact mode.

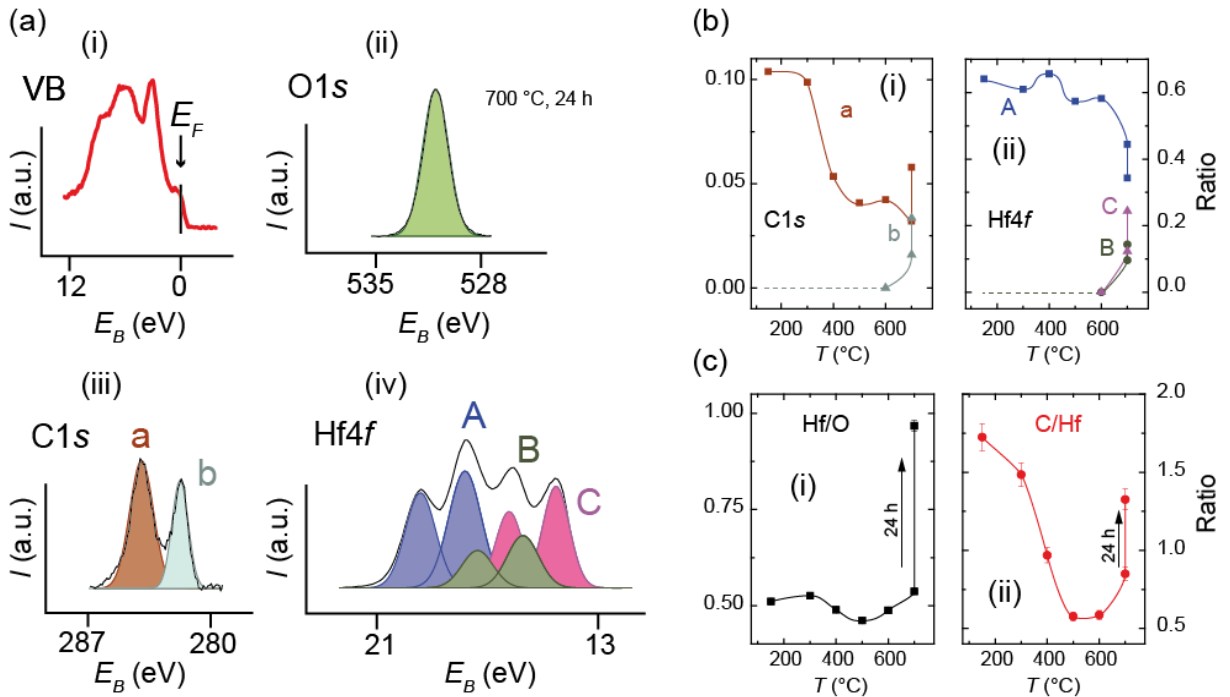

Fig. 2: Influence of the UHV reduction process on a highly oxygen deficient HfO$_{2-x}$ thin film (grown under 0.3 sccm oxygen flow rate on $c$-cut sapphire). (a) Results after annealing at 700 °C for 24 h (maximal reduction): (i) Valence band spectrum; (ii) O1$s$ core spectrum; (iii) C1$s$ core spectrum; (iv) Hf4$f$ spectrum. (b) Evolution of peak intensities during the stepwise annealing as obtained by peak simulation: (i) C1$s$ core level; (ii) Hf4$f$ spectra. (c) Evolution of elemental ratios during the stepwise annealing: (i) Hf/O ratio; (ii) C/Hf ratio.

At first, we have quantified the amount of carbon in our MBE grown films by Tof-SIMS measurements on one hafnium oxide thin film. For absolute quantification, an additional reference sample has been fabricated by implanting a well-defined amount of carbon into the film using an implanter.[15,16] Fig. 1 shows the depth profile of a 100 nm thick oxygen deficient $HfO_{2-x}$ thin film deposited on *c*-cut sapphire. Based on the nearly constant intensity for the carbon C$^-$ signal between 20 and 80 nm film depth, a carbon concentration of $2\cdot 10^{19}$ /cm³ was determined. This concentration is comparably low to known carbon concentrations for ALD-grown films, which are in the range of $7\cdot 10^{19}$ /cm³.[17] The source of carbon impurities in the here described experiments is residual graphite from the UHV chamber due to the use of carbon spray. The Hf source itself does not contain carbon contaminations as ruled out by SIMS studies.

XPS measurements of the core lines O1*s*, C1*s*, and Hf4*f* were performed on stoichiometric and oxygen deficient thin films as a function of temperature. For this purpose, the *ex situ* transferred samples were subjected to a temperature treatment from room temperature up to 700 °C in steps of 100 °C inside the XPS chamber. The dwell time for each annealing step was 1 h, whereas at 700 °C two measurements were performed, one after 1 h and one after 24 hours dwell time. For annealing temperatures below 700 °C, the core lines of O1*s* (531 eV) and C1*s* (285 eV) could be refined by one component, whereas the Hf4*f* core line was fitted by a single doublet as the energy levels for Hf4*f* are split off by 1.7 eV due to spin orbit coupling (225 and 214 eV, not shown).[18] The position of this doublet is sensitive to the Hf-O binding in $HfO_2$. For all spectra, charging effects were subtracted before refinement. The observation of carbon on the sample surface for *ex situ* samples is expected, as during handling under atmospheric conditions carbon-based molecules adsorb on the film surface. The observed carbon peak position corresponds to C-C and C-H bonds detected for both, stoichiometric and oxygen deficient films.

First, we describe the XPS spectra for oxygen deficient $HfO_{2-x}$ thin films (grown under 0.3 sccm oxygen flow rate) after annealing at 700 °C for 24 hours as shown in Fig. 2(a). The valence band spectra (Fig. 2(a)(i)) show conducting states at the Fermi level indicating the formation of a conducting surface layer. One origin of this layer is the extremely oxygen deficient layer itself as shown previously by XPS and hard X-ray photoelectron spectroscopy.[9,10] In this case, the conductivity arises from overlapping oxygen defect states.[7] The second origin is the formation of $HfC_x$, i.e. Hf-C bonds near the surface. The valence band spectra can then be interpreted as hybridized C-*p* and Hf-*d* states.[19] There is no change in the O1*s* core line as compared to lower temperatures. It can always be refined by one peak originating from Hf-O bonds (see Fig. 2(a)(ii)).

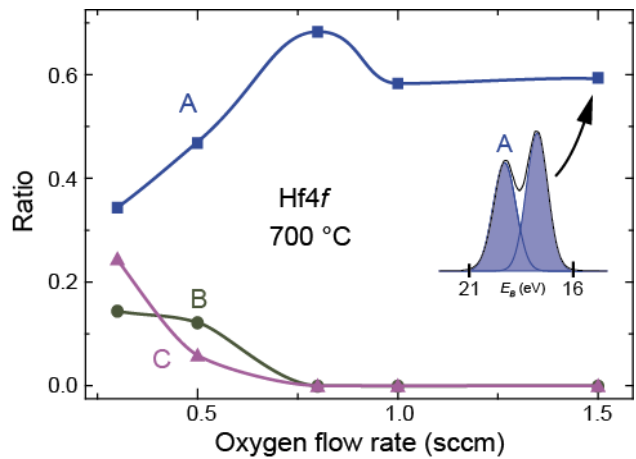

Fig. 3: Peak ratios of the simulated peaks of the Hf4*f* core line measured by XPS after 24 h annealing at 700 °C on films grown under different oxygen flow rates. The inset shows the Hf4*f* spectrum after annealing for the fully stoichiometric films grown under 1.5 sccm oxygen flow.

The most important difference to stoichiometric films occurs in the C1*s* and Hf4*f* spectra (see Fig. 2(a)(iii) and (iv)). At First, considering carbon, the peak labeled *a* at 284 eV is due to the known surface bonds C-C/C-H, whereas the second peak *b* at 282 eV can be attributed to Hf-C formation. This is supported by the Hf4*f* spectrum where two new doublets appear labeled *B* and *C* (see Fig. 2(a)(iv)). The carbon peak *b* and the Hf doublets *B* and *C* only occur for highly oxygen deficient films (oxygen flow rate equal or below 0.5 sccm), they are not observed for stoichiometric films which have been treated in exactly the same manner. The Hf doublet *B* indicates an intermediate valence state of Hf due to the high amount of oxygen vacancies which needs to be compensated. The doublet *C* is close to the position of metallic Hf. However, the Hf spectrum in HfC shows the Hf doublet at the same position due to the metallic bonding character of the compound.[20,21] Together with the rather unambiguous signal from the carbon line *b*, we interpret the Hf4*f* doublet *C* as indicative for the formation of $HfC_x$.

Fig. 2 (b)(i) and (ii) shows the variation of the components *a* and *b* of the C1*s* signal as well as components *A*, *B*, and *C* of Hf4*f* as a function of annealing temperature. Elemental ratios were obtained by weighting the spectral intensities using the specific sensitivity factors as implemented in the Multipak software. All displayed ratios were normalized to the full integral specular intensity of the XPS spectrum (C1*s*, Hf4*f*, O1*s*). Peak *b* is detectable for the first time at 600 °C. When Hf is placed in a strongly oxygen deficient environment with dangling bonds, the formation of HfC seems to be facilitated. The formation temperature of HfC which is of the order 2000 °C is substantially reduced. The doublets *B* and *C* from the Hf4*f* spectrum exactly show the same temperature evolution, supporting the HfC formation. This evolution describes a change in bonding from a covalent-ionic bonding between Hf and O to a covalent-metallic bonding between Hf and adventitious C, suggesting the formation of hafnium carbide.[22,23] The spectral intensity of component *b* in the C1*s* spectra decreases with increasing XPS analyzer angle. This observation suggests that the formation of C-Hf bonds is only taking place at the film surface.

The formation of hafnium carbide is additionally supported by the elemental ratios shown in Fig 2 (c)(i) and (ii). The Hf/O ratio does not change significantly for annealing temperatures below 700 °C, but strongly increases for 700 °C. This indicates the decomposition of the oxidized surface $HfO_2$. The C/Hf ratio decreases with increasing temperature up to 600 °C due to desorption of adsorbed carbon/carbon compounds on the surface. At 700 °C, the C/Hf ratio increases again, as the activated surface starts to incorporate carbon forming hafnium carbide. Comparable studies on titanium oxide also show carbide formation above 1100 °C for nanopowders and above 475 °C for thin films.[24]

When the same annealing experiment is repeated for fully oxidized $HfO_2$ samples (1.5 sccm oxygen flow during growth), even after 24 h annealing time, only the doublet *A* is observed in the XPS Hf4*f* spectrum (see inset of Fig. 3). This clearly shows that Hf-C bonds only form under strongly oxygen deficient conditions. We have measured in total 5 thin film samples with different oxygen flow rates during MBE growth. The main panel of Fig. 3 shows the consistent evolution of the observed XPS spectra with increasing oxygen deficiency. The Hf4*f* doublets *B* and *C* are not observed for oxygen flows above 0.5 sccm. For these samples only doublet *A* belonging to $Hf^{4+}$ is observed in the XPS spectra.

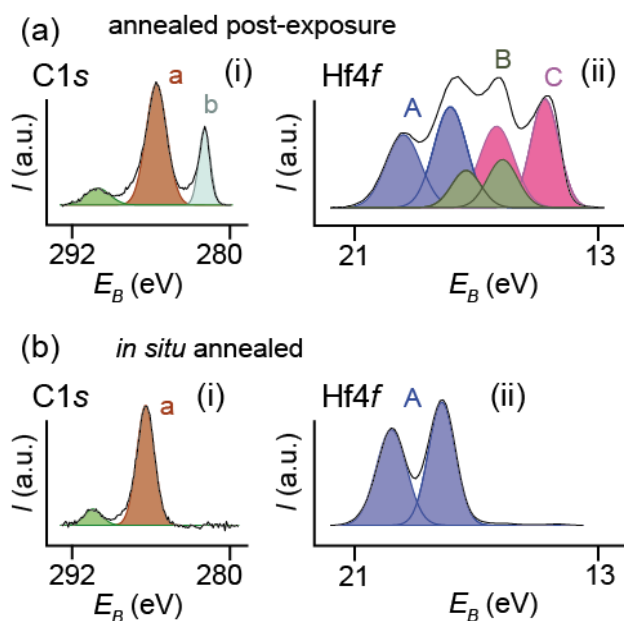

Fig. 4: XPS spectra of the Hf4*f* and C1*s* peak for vacuum annealed $HfO_{2-x}$ samples (6 hours at 700 °C). (a) These spectra are taken from a sample which has been first exposed to air, and subsequently annealed. (i) C1*s*, and (ii) Hf4*f* spectrum. (b) These spectra are taken from a sample which was annealed *in situ* (without air exposure): (i) C1*s*, and (ii) Hf4*f* spectrum.

The most important result of the present study is that oxygen deficiency acts as a catalytic converter for the formation of HfC at unexpectedly low temperatures. Since conducting filaments in RRAM devices are modeled as aggregation of oxygen vacancies, this process might occur in such devices due to Joule heating which easily can produce temperature of several 100 degrees given the nano diameter of the filaments. In our experiments, we have two possible sources of carbon, the incorporated carbon during thin film growth with a concentration of $2 \cdot 10^{19}$ /cm³, or adsorbed carbon-based molecules due to the *ex situ* processing of our samples. In order to distinguish between these two sources, we have performed the following experiments: Two films of highly deficient hafnium oxide were synthesized in the same deposition. The first sample was *in situ* annealed for 6 h at 700 °C. The second sample was exposed to atmospheric conditions and then annealed under the same conditions.

We show the room-temperature XPS spectra for the C1$s$ and Hf4$f$ lines of these samples in Fig. 4 (a) and (b). The clear result is that only the sample that had been exposed to air shows the formation of HfC$_x$ as identified from the additional peak *b* in the C1s spectrum (see Fig. 4(a)(i)) and the additional Hf4$f$ doublets *B* and *C* (see Fig. 4(a)(ii)). These peaks are not detected for the sample that had not been exposed to air (see Fig. 4(b)(i) and (ii)). Since only exposed film shows stable Hf-C bonds, the source of carbon can only be the adsorbed carbon species during the exposure. The small peak in the C1s spectra at ~290 eV observed for both films is indicative for C-O bonds due to the *ex situ* transport to the XPS chamber. Our result suggests that a carbon concentration of $2 \cdot 10^{19}$ /cm³ is not sufficient to provide any measurable Hf-C layer at the sample surface. The exposure to air with subsequent annealing, however, provides sufficient carbon species for a large spectral weight of HfC$_x$. While in our case, both sources of carbon could be easily avoided in a more advanced device fabrication process, other deposition methods used in industrial processes such as ALD may lead to higher concentration levels of carbon. Although the thermal budget in standard CMOS processing does not reach temperatures of 700 °C, it has been calculated that during resistive switching Joule heating occurs with temperatures of more than 1000 K.[25] This could lead to a local carbide formation within the conductive filament itself influencing its switching properties.

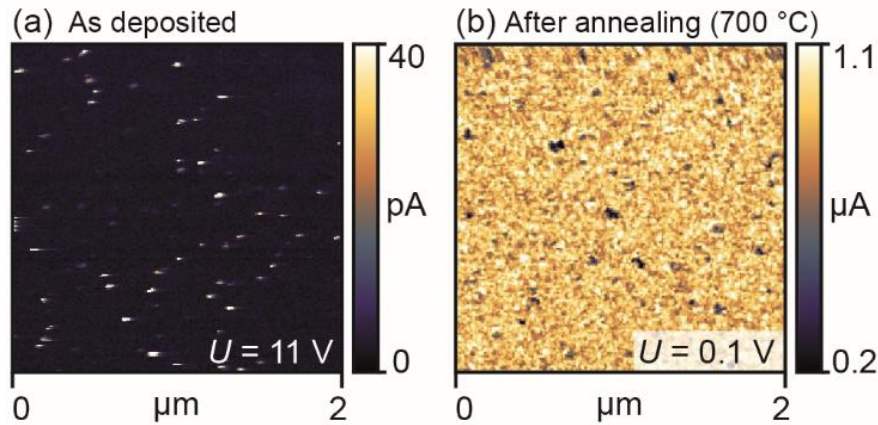

Fig. 5: LC-AFM measurements on the highly oxygen deficient film grown under 0.3 sccm oxygen before (a) and after (b) annealing under UHV conditions at 700 °C.

In order to better understand the influence of HfC$_x$ on electrical transport and resistive switching, we have performed LC-AFM measurements *ex situ* on the film grown under 0.3 sccm oxygen flow rate before and after annealing in reducing atmosphere (24 h, 700 °C). The LC-AFM method is a highly surface sensitive tool allowing electrical characterization of only the top section of the investigated films, which is in the case of the annealed film hafnium carbide or a mixture of hafnium carbide, HfO$_2$, and HfO$_{2-x}$ as concluded from the XPS results. In Fig. 5(a) and (b) the current maps are shown obtained while scanning an area of 2 x 2 µm². For the surface of the film before annealing (Fig. 5(a)) the conductivity was very low and we had to choose a voltage applied to the tip of +11 V in order to measure current. It can be seen that also under this high forward bias only a few conducting areas were present. This reveals that a thin insulating layer has formed during the exposure to ambient conditions due to carbon contamination and oxidation into an HfO$_2$ surface layer preventing an electrical connection from the AFM tip to the interior of the film. As demonstrated in our previous study[9], the as-deposited oxygen-deficient film shows however much higher conductivity and controllable resistive switching when contacted with deposited metal electrodes percolating the surface layer. In contrast to the as-deposited film, the LC-AFM current maps of the vacuum-annealed film (Fig. 5(b)) reveal a high surface conductivity even at a moderate reading voltage of +0.1 V reflecting the formation of stable metallic hafnium carbide with a grainy microstructure showing conducting clusters with slight variations in local conductivity and diameters in the range of 20-50 nm. Since HfC is known to be a very stable material that is not expected to dissolve again by electrical stimuli, the reset into the off state may be prevented as observed in carbon containing devices[11]. In conclusion, the LC-AFM analysis reveals that the surface properties with respect to electronic transport of the oxygen deficient films can be manipulated significantly in gradients of the chemical potential by formation of HfC.

In summary, we have shown that hafnium carbide forms in strongly oxygen deficient hafnium oxide thin films at temperatures of several 100 degrees, well below its thermodynamic equilibrium formation temperature. The evolution of hafnium carbide increases the surface conductivity significantly. A sufficient source of carbon impurities for carbide formation under the described conditions at the sample surface are adsorbed carbon species during *ex situ* processes. Given the large amount of oxygen vacancies inside conducting filaments and the high temperatures due to Joule heating in the electroforming and resistive switching process, the role of carbon impurities in general is important for RRAM devices. In particular, for HfO$_{2-x}$ (or similar materials) based devices the formation of HfC has to be taken into account when a sufficient amount of carbon is available.

We thank A. Besmehn for XPS measurements and A. Dahmen for carbon implantation. We gratefully acknowledge G. Cherkashinin for fruitful discussions. This work was supported in part by the Deutsche Forschungsgemeinschaft under projects SFB 917 and AL560/13-2. Funding by the Federal Ministry of Education and Research (BMBF) under contract 16ES0250 is also gratefully acknowledged. We thank funding by ENIAC JU within the project PANACHE.

[1] J.H. Choi, Y. Mao, J. and P. Chang, Mater. Sci. Eng. R-Rep. **72**, 97136 (2011).

[2] R. Waser, R. Dittmann, G. Staikov, and K. Szot, Adv. Mater. **21**, 2632 (2009).

[3] C. Walczyk, C. Wenger, R. Sohal, M. Lukosius, A. Fox, J. Dabrowski, D. Wolanski, B. Tillack, H.-J. Muessig, and T. Schroeder, J. Appl. Phys. **105**, 114103 (2009).

[4] P. Gonon, M. Mougenot, C. Vallee, C. Jorel, V. Jousseaume, H. Grampeix, and F. El Kamel, J. Appl. Phys. **107**, 74507 (2010).

[5] G. Bersuker, J. Yum, L. Vandelli, A. Padovani, L. Larcher, V. Iglesias, M. Porti, M. Nafria, K. McKenna, A. Shluger, P. Kirsch, and R. Jammy, Solid State Electron. **65-66**, 146 (2011).

[6] M. Lanza, G. Bersuker, M. Porti, E. Miranda, M. Nafria, and X. Aymerich, Appl. Phys. Lett. **101**, 193502 (2012).

[7] E. Hildebrandt, J. Kurian, M. M. Mueller, T. Schroeder, H. J. Kleebe, and L. Alff, Appl. Phys. Lett. **99**, 112902 (2011).

[8] E. Hildebrandt, J. Kurian, L. Alff, J. Appl. Phys. **112**, 114112 (2012).

[9] S. U. Sharath, T. Bertaud, J. Kurian, E. Hildebrandt, C. Walczyk, P. Calka, P. Zaumseil, M. Sowinska, D. Walczyk, A. Gloskovskii, T. Schroeder, and L. Alff, Appl. Phys. Lett. **104**, 063502 (2014).

[10] S. U. Sharath, J. Kurian, P. Komissinskiy, E. Hildebrandt, T. Bertaud, C. Walczyk, P. Calka, T. Schroeder, and L. Alff, Appl. Phys. Lett. **105**, 073505 (2014).

[11] M. Sowinska, T. Bertaud, D. Walczyk, S. Thiess, P. Calka, L. Alff, C. Walczyk, and T. Schroeder, J. Appl. Phys. **115**, 204509 (2014).

[12] J. Celinska; C. McWilliams; C. A. P. de Araujo; K. H. Xue, Journal of Applied Physics 109, 091603 (2011)

[13] K. Szot, W. Speier, G. Bihlmayer, and R. Waser, Nat. Mater. **5**, 312 (2006).

[14] C. Rodenbücher, W. Speier, G. Bihlmayer, U. Breuer, R. Waser, and K. Szot, New J. Phys. **15**, 103017 (2013).

[15] J. F. Gibbons, Proc. IEEE **56**, 295 (1968).

[16] R. P. Gittins, D.V. Morgan, and G. Dearnaley J. Phys. D **5**, 1654 (1972).

[17] S. Kamiyama, T. Miura, and Y. Nara, Appl. Phys. Lett. **87**, 132904 (2005).

[18] C. Morant, L. Galán, and J. M. Sanz, Surf. Interface Anal. **16**, 304 (1990).

[19] Q. Zeng, J. Peng, A. R. Oganov, Q. Zhu, C. Xie, X. Zhang, D. Dong, L. Zhang, and L. Cheng, Phys. Rev. B **88**, 214107 (2013).

[20] A.A. Lavrentyev, B.V. Gabrelian, V.B. Vorzhev, I.Ya. Nikiforov, O.Yu. Khyzhunb, and J.J. Rehr. J. Alloys Compd. **462**, 4 (2008).

[21] G. R. Gruzalski and D. M. Zehner. Phys. Rev. B **42**, 2768 (1990).

[22] P. Weinberger, R. Podloucky, C. P. Mallett, and A. Neckel, J. Phys. C: Solid State Phys. **12**, 801 (1979).

[23] Q. Zeng, J. Peng, A. R. Oganov, Q. Zhu, C. Xie, X. Zhang, D. Dong, L. Zhang, and L. Cheng, Phys. Rev. B **88**, 214107 (2013).

[24] L. Calvillo, D. Fittipaldi, C. Rüdiger, S. Agnoli, M. Favaro, C. Valero-Vidal, C. Di Valentin, A. Vittadini, N. Bozzolo, S. Jacomet, L. Gregoratti, J. Kunze-Liebhäuser, G. Pacchioni and G. Granozzi, J. Phys. Chem. C **118**, 22601 (2014).

[25] E. Yalon, S. Cohen, A. Gavrilov and D. Ritter, Nanotechnology **23**, 465201 (2012).